
\input lanlmac


\def\th{\theta}

\def\cob{\delta}
\def\ep{\epsilon}

\def\D{{\bf D}}

\def\hf{{1\over 2}}

\def\o{\over}

\def\til#1{\widetilde{#1}}

\def\si{\sigma}

\def\del{\partial}

\def\bra{\langle}
\def\ket{\rangle}
\def\lf{\left}
\def\ri{\right}
\def\riya{\rightarrow}

\def\la{\lambda}
\def\La{\Lambda}
\def\h#1{\widehat{#1}}

\def\om{\omega}

\def\st{\star}
\def\A{{\cal A}}
\def\D{{\cal D}}
\def\F{{\cal F}}

\def\np#1#2#3{{ Nucl. Phys.} {\bf B#1}, #2 (19#3)}

\def\pln#1#2#3{{Phys. Lett.} {\bf B#1}, #2 (19#3)}

\def\pr#1#2#3{{ Phys. Rev.} {\bf D#1}, #2 (19#3)}

\def\hpt#1{{\tt hep-th/#1}}

\lref\SW{
N. Seiberg and E. Witten,
``String Theory and Noncommutative Geometry'',
JHEP {\bf 9909}, 032 (1999), \hpt{9908142}.
}
\lref\ACNY{
A. Abouelsaood, C. G. Callan, C. R. Nappi and S. A. Yost,
``Open Strings in Background Gauge Fields'',
\np{280}{599-624}{87}.
}
\lref\barbon{
E. Alvarez and J. L. F. Barb\'{o}n and J. Borlaf,
``T-duality for Open Strings'',
\np{479}{218-242}{96}, \hpt{9603089}.
}
\lref\chuho{
C.-S. Chu and P.-M. Ho,
``Noncommutative Open String and D-brane'',
\np{550}{151-168}{99}, \hpt{9812219};
``Constrained Quantization of Open String in Background B Field 
and Noncommutative D-brane'', \hpt{9906192}.
}
\lref\KO{
T. Kawano and K. Okuyama,
``Matrix Theory on Noncommutative Torus'',
\pln{433}{29-34}{98}, \hpt{9803044}.
}
\lref\Ishin{
N. Ishibashi,
``A Relation between Commutative and Noncommutative Descriptions 
of D-branes'', \hpt{9909176}.
}
\lref\Ishio{
N. Ishibashi,
``$p$-branes from $(p-2)$-branes in the Bosonic String Theory'',
\np{539}{107-120}{99}, \hpt{9804163}.
}
\lref\Kont{
M. Kontsevich,
``Deformation quantization of Poisson manifolds'',
{\tt q-alg/9709040}.
}
\lref\CF{
A. S. Cattaneo and G. Felder,
``A path integral approach to the Kontsevich quantization formula'',
{\tt math.QA/9902090}.
}
\lref\corn{
L. Cornalba and R. Schiappa,
``Matrix Theory Star Products from the Born-Infeld Action'',
\hpt{9907211}\semi
L. Cornalba,
`` D-brane Physics and Noncommutative Yang-Mills Theory'',
\hpt{9909081}.
}
\lref\callan{
C. G. Callan, C. Lovelace, C. R. Nappi and S. A. Yost,
``Loop Corrections to Superstring Equations of Motion'',
\np{308}{221-281}{88}. 
}

\lref\branes{
T. Banks, N. Seiberg and S. Shenker,
``Branes from Matrices'',
\np{490}{91-106}{97}, \hpt{9612157}.
}
\lref\BFFS{
T. Banks, W. Fischler, S. H. Shenker and L. Susskind,
``M Theory As A Matrix Model: A Conjecture'',
\pr{55}{5112-5128}{97}, \hpt{9610043}.
}
\lref\town{
P. K. Townsend,
``D-branes from M-branes'',
\pln{373}{68-75}{96}, \hpt{9512062}.
}
\lref\jab{
F. Ardalan, H. Arfaei and M. M. Sheikh-Jabbari,
`` Noncommutative Geometry From Strings and Branes'',
JHEP {\bf 9902}, 016 (1999), \hpt{9810072};
``Dirac Quantization of Open Strings and Noncommutativity in Branes'',
\hpt{9906161}\semi
M. M. Sheikh-Jabbari,
``Open Strings in a B-field Background as Electric Dipoles'',
\pln{455}{129-134}{99}, \hpt{9901080}.
}
\lref\scho{
V. Schomerus,
`` D-branes and Deformation Quantization'',
JHEP {\bf 9906}, 030 (1999), \hpt{9903205}.
}
\lref\miao{
M. Li,
``Strings from IIB Matrices'',
\np{499}{149-158}{97}, \hpt{9612222}.
}
\lref\CDS{
A. Connes, M. R. Douglas and A. Schwarz,
``Noncommutative Geometry and Matrix Theory: 
Compactification on Tori'',
JHEP {\bf 9802}, 003 (1998), \hpt{9711162}.
}
\lref\DH{
M. R. Douglas and C. Hull,
``D-branes and the Noncommutative Torus'',
JHEP {\bf 9802}, 008 (1998), \hpt{9711165}.
}
\lref\KO{
T. Kawano and K. Okuyama,
``Matrix Theory on Noncommutative Torus'',
\pln{433}{29-34}{98}, \hpt{9803044}.
}
\lref\cheung{
Y. E. Cheung and M. Krogh, 
``Noncommutative Geometry from D0-branes in a Background B-field'',
\np{528}{185-196}{98}, \hpt{9803031}.
}
\lref\HWW{
P.-M. Ho, Y.-Y. Wu and Y.-S. Wu,
``Towards a Noncommutative Geometric Approach to Matrix
Compactification'',
\hpt{9712201}\semi
P.-M. Ho and Y.-S. Wu,
``Noncommutative Gauge Theories in Matrix Theory'',
\hpt{9801147}.
}
\lref\Garo{
M. R. Garousi,
``Non-commutative world-volume interactions on D-brane and Dirac-Born-Infeld
action'', Nucl. Phys. {\bf B579}  209-228 (2000), \hpt{9909214}; 
``Tachyon couplings on non-BPS D-branes and Dirac-Born-Infeld action'',
\hpt{0003122}.
}

\Title{             
                                             \vbox{\hbox{KEK-TH-655}
                                             \hbox{hep-th/9910138}}}
{\vbox{\centerline{A Path Integral Representation of the Map between}
\vskip .2in
\centerline{ Commutative and Noncommutative Gauge Fields}}}

\vskip .2in

\centerline{
                    Kazumi Okuyama
}

\vskip .2in

\centerline{\sl High Energy Accelerator Research Organization (KEK)}
\centerline{\sl Tsukuba, Ibaraki 305-0801, Japan}
\centerline{\tt kazumi@post.kek.jp }

\vskip 3cm
\noindent

The world-volume theory on a $D$-brane in a constant
$B$-field background
can be described by either
commutative or noncommutative Yang-Mills theories.
These two descriptions correspond to
two different gauge fixing of the diffeomorphism on
the brane.  
Comparing the boundary states in the two gauges, we derive a map between
commutative and noncommutative gauge fields in a path integral form,
when the gauge group is $U(1)$.

\Date{October, 1999}

\vfill
\vfill

\newsec{Introduction}
Noncommutativity of coordinates appears in the study of $D$-branes
in two apparently different situations.
One such situation occurs when $N$ $D$-branes coincide.
Then their transverse coordinates are promoted to $N\times N$ matrices.
Another situation is the one in which
the boundary coordinates  $X^i(\tau)$ 
of an open string become
noncommutative in the presence of a constant NS-NS $B$-field
\refs{\ACNY,\chuho,\jab,\scho}. The commutation relations of
$X^i(\tau)$'s are written as
\eqn\comXth{
[X^i(\tau),X^j(\tau)]=i\th^{ij}.
}
These relations lead to the noncommutativity 
of the world-volume coordinates of $D$-branes,
which was first appeared in the compactification of Matrix theory
in a three-form field background \refs{\CDS\DH\KO\cheung{--}\HWW}. 

As was pointed out in \Ishio, in a sense
these two noncommutativity are ``dual'' to
each other. A $D$-brane in a constant $B$-field background
can be described as
a collection of infinitely many lower dimensional $D$-branes 
\refs{\Ishio,\town,\BFFS,\branes}.
In this description, the transverse coordinates of lower dimensional 
$D$-branes should satisfy the same relation as \comXth.

In \SW, Seiberg and Witten argued that the theory on $D$-branes in 
a $B$-field background can be described by either commutative or
noncommutative Yang-Mills theories and these descriptions 
correspond to Pauli-Villars and  
point splitting regularizations of the world-sheet theory, respectively.
They derived the relation between the commutative gauge field $A$
and the noncommutative gauge field $\h{A}$ by requiring the equivalence of
the gauge transformation of $A$ and $\h{A}$.

In the $D$-brane world-volume perspective, these two descriptions 
correspond to two different 
gauge fixing of the world-volume diffeomorphism \refs{\Ishin,\corn}.
One is the static gauge and the other is 
the ``constant field strength gauge'' 
(in the following, we will call the
latter gauge ``$\F=\om$ gauge''). In the static gauge, 
the coordinates parallel to the brane are fixed and the ordinary 
gauge field remains as a dynamical field. In $\F=\om$ gauge,
the fluctuation of ordinary gauge field is set to zero.
In this gauge, the
dynamical degree of freedom are carried by the scalar fields
corresponding to the parallel coordinates of the brane. 
The noncommutative gauge field appears as the fluctuation of this scalar
field around the static gauge configuration. 
The two different descriptions are mapped to each other by the
world-volume diffeomorphism.

In \refs{\Ishin,\corn}, the relation between $A$ and $\h{A}$ 
is derived from the diffeomorphism invariance of $D$-branes,
but only in a semiclassical
sense, i.e. the Moyal bracket is replaced with the Poisson bracket.
To ``quantize'' the Poisson bracket into Moyal bracket, it is useful
to use the path integral formalism \refs{\Kont, \CF}.
In this paper, we will derive the map 
$A \mapsto \h{A}$ in a path integral form by comparing the 
boundary states in the two different gauges, when the gauge group is $U(1)$.
The main claim of this paper is that the map is given by
\eqn\main{\eqalign{
&\int \D\xi(\si) \,\exp\lf(i\int d\si \,\hf\om_{ij}\xi^i\del_{\si}\xi^j
+A_i(y+\xi)\del_{\si}(y^i+\xi^i)\ri) \cr
=&\int \D\xi(\si) \,\exp\lf(i\int d\si \,\hf\om_{ij}\xi^i\del_{\si}\xi^j
+\h{A}_i(y+\xi)\del_{\si}y^i\ri)
}}
where $\om_{ij}=(\th^{-1})_{ij}$ and $y^i(\si)$ is an arbitrary function.
We will show that 
this relation satisfies the requirement for the mapping between
$A$ and $\h{A}$, namely the equivalence of the gauge transformations for
$A$  and $\h{A}$. 

Here we comment on the relation between \main\ and the 
interpretation of the origin of noncommutativity in \SW.   
The left hand side of \main\ 
is divergent due to the contraction of $\xi(\si)$ and $\del_{\si}\xi(\si)$
at the same $\si$. We regularized it preserving the ordinary gauge
invariance of $A$. On the other hand,
the right hand side is finite because of the absence of this
contraction. Therefore \main\ naturally realizes the idea of \SW,
namely the commutative and noncommutative descriptions of 
gauge theory on $D$-branes correspond to the two different regularizations
(point-splitting and Pauli-Villars) of the worldsheet theory.

This paper is organized as follows: In section 2, we review the
symmetries of  boundary states and the derivation of the map in 
the semiclassical form. In section 3, we derive the relation \main\
by comparing the boundary states in the two different gauges and show that
it gives the correct relation between $A$ and $\h{A}$. We also argue how to 
regularize the left hand side of \main\ without breaking the ordinary
gauge invariance of $A$.   
Section 4 is devoted to discussions.

\newsec{Symmetries of $D$-brane Boundary States}
In this section, we consider boundary states in a constant 
background $B$-field and their symmetries. 
In the following, we will gauge away the $B$-field in the bulk
of the world-sheet and treat it as a gauge field background
 with constant field strength.
To construct boundary states, it is convenient to 
introduce the coherent state $|x\ket$ defined by 
$X^i(\si)|x\ket=x^i(\si)|x\ket$. $|x\ket$ can be  written as 
\eqn\cohbyp{
|x\ket=\exp\lf(-i\int d\si P_i(\si)x^i(\si)\ri)|D\ket
}
where $|D\ket$ is the Dirichlet boundary state defined by
$X^i(\si)|D\ket=0$ and $P_i$ is the
momentum conjugate to $X^i$.  When we consider a $Dp$-brane, $i$
runs from 0 to $p$. 
Using $|x\ket$, the boundary state coupled to a $U(1)$ gauge field $\A$
is given by \callan
\eqn\BtodelS{
|B\ket=\int \D x e^{i\int \A(x)}|x\ket.
}
In the case of the constant field strength $\F_{ij}=\om_{ij}$,  
\BtodelS\ is reduced to
\eqn\onshellB{
|B\ket=\int \D x \exp\lf(i\int d\si \hf\om_{ij}x^i\del_{\si}x^j
-P_ix^i\ri)|D\ket.
}
In the following, we assume that $p$ is odd and $\om_{ij}$ is invertible. 

Now we consider the fluctuation of the $Dp$-brane around the above
configuration.
The most general form of the boundary state is
\eqn\genBAphi{
|\A,\phi\ket=\int \D\phi(x)\exp\lf(i\int d\si 
\big(\A_i(x)\del_{\si}x^i-P_i\phi^i(x)\big)\ri)|D\ket,
}
where $\A_i$ is the gauge field 
on the $Dp$-brane and $\phi^i$ is the scalar field 
corresponding to the coordinate parallel to the $Dp$-brane. 
We suppress the transverse coordinates of the brane for simplicity. 
The fluctuation around the configuration \onshellB\ is parameterized as
\eqn\defAa{
\A_i=\hf\om_{ji}x^j+A_i,~~\phi^i=x^i+\th^{ij}a_j.
}

Following \refs{\Ishin,\corn}, we review the argument that $A_i$ and 
$a_i$ become the ordinary and noncommutative gauge fields, respectively,
after the gauge fixing of the diffeomorphism on the $D$-brane.
So as to make $|\A,\phi\ket$ diffeomorphism invariant,
we choose the measure $\D\phi(x)$ in \genBAphi\ as
\eqn\mesDphi{
\D\phi(x)=\prod_{\si}dx(\si)\det\lf({\del \phi^i\o\del x^j}\ri)
\big(x(\si)\big).
}
Under the diffeomorphism on the $Dp$-brane, $\A$ and $\phi^i$ transform 
as a 1-form and a scalar, respectively, i.e. 
\eqn\trfAphi{\eqalign{
&\cob_{{\rm diff}} \A={\cal L}_v\A=(di_v+i_vd)\A, \cr
&\cob_{{\rm diff}}\phi^i={\cal L}_v\phi^i=v^k\del_k\phi^i.
}}
$|\A,\phi\ket$ is also invariant under the gauge transformation 
\eqn\gaugetrf{
\cob_{{\rm gauge}} \A=d\ep,~~\cob_{{\rm gauge}}\phi=0,
}
and the canonical transformation
\eqn\canFtrf{
\cob_{{\rm can}} \A=0,~~\cob_{{\rm can}} \phi={\cal L}_{{\rm ham}\la}\phi,
}
where ${\rm ham}\la$ is a Hamiltonian vector field 
defined by
\eqn\defsgf{
i_{{\rm ham}\la}\F=d\la.
}
We can see that the canonical transformation 
is equivalent to the field dependent gauge transformation
up to the diffeomorphism:
\eqn\uptodiffeq{
\cob_{{\rm can}}(\la)=-\cob_{{\rm gauge}}(\la+i_{{\rm ham}\la}\A)
+\cob_{{\rm diff}}({\rm ham}\la).
}

By fixing the diffeomorphism invariance, we can 
obtain two different pictures for the same state.
The first is the ``static gauge'' which is defined by $\phi^i=x^i$.
In this gauge, $|\A,\phi\ket$ is reduced to
\eqn\staticB{
|\A\ket=\int \D xe^{i\int \A-P_ix^i}|D\ket.
}
The second is the ``$\F=\om$ gauge''.
In this gauge, the fluctuation of
$\A$ is set to be zero, and $|\A,\phi\ket$ becomes
\eqn\sheetB{
|\phi\ket=\int \D\phi(x)e^{i\int \hf\om_{ij}x^idx^j-P_i\phi^i}|D\ket.
}
For $|\A\ket$, the residual diffeomorphism invariance is the canonical
transformation with respect to the symplectic form $\F$
which coincides with the
usual gauge symmetry for $\A$, or for $A$. On the other hand,
$|\phi\ket$ has no gauge field but it has a residual diffeomorphism 
symmetry which preserves the symplectic form $\om$.
Its action on $\phi$ is given by 
\eqn\cantrfonphi{
\cob \phi=\{\phi,\la\}=\th^{kl}\del_k\phi\del_l\la.
}
In terms of $a_i$, this symmetry is written as 
$\th^{ij}\cob a_j=\{\phi^i,\la\}$, or
\eqn\trfa{
\cob a_i=\del_i\la+\th^{kl}\del_ka_i\del_l\la.
}

Since the two states $|\A\ket$  and $|\phi\ket$ 
correspond to two different gauge choice for the same state, they
should be equivalent under the diffeomorphism. 
Under the change of variable 
\eqn\chvarpath{
x=\phi(y)
}
in the path integral  \staticB,
the equivalence of the two states \staticB\ and \sheetB\ requires
\eqn\Ishtrf{
\phi^*\A=\hf\om_{ij}y^idy^j+d(*).
}
This relation gives a nontrivial mapping between $A_i$ and $a_i$ 
\refs{\Ishin,\corn}.

Although one can show that $a_i$ is equal to the noncommutative
gauge field $\h{A}$ up to the second
order in $\th$-expansion, $a_i$ is obviously not equal to $\h{A}$ since
the gauge transformation
 for $a_i$ \trfa\ is given in terms of the Poisson bracket 
instead of the Moyal bracket.

\newsec{The Map between Commutative and Noncommutative Gauge Fields}
In this section, we propose the resolution of the discrepancy
between $a_i$ and $\h{A}_i$ and give a simple rule for 
the map between $A_i$ and $\h{A}_i$. 
As we will see below, 
in order to realize the noncommutative gauge symmetry for $\h{A}$,
we should change the integration measure in $|\phi\ket$ from 
$\D\phi(x)$ to the flat measure $\D x$.
We write the state with the measure $\D x$ as   
\eqn\flatphi{
|\phi\ket_{{\rm NC}}=\int\D x\exp\lf(i\int d\si \hf\om_{ij}x^i\del_{\si}x^j
-P_i\phi^i(x)\ri)|D\ket.
}
With this measure, the fluctuation of $\phi^i$ can be identified with
$\h{A}$: 
\eqn\phitohA{
\phi^i(x)=x^i+\th^{ij}\h{A}_j(x).
}

To show that $|\phi\ket_{{\rm NC}}$ is invariant under 
the noncommutative gauge transformation for $\h{A}$,
it is convenient to use the T-dual picture.  
In terms of the coherent state $\til{|y\ket}$ 
for the T-dual coordinate, $|D\ket$ is written as
\eqn\DtilN{
|D\ket=\int \D y\til{|y\ket}.
}
On $\til{|y\ket}$, $P_i$ acts as $\del_{\si}y_i$.
Therefore, the original  and the T-dual coherent
state are related \barbon\ by
\eqn\cohinT{
|x\ket=\int \D y \exp\lf(-i\int d\si\del_{\si}y_ix^i\ri)\til{|y\ket}.
}
Using these relations, $|\phi\ket_{{\rm NC}}$  can be written as
\eqn\phiinxi{\eqalign{
|\phi\ket_{{\rm NC}}&=\int\D x\D y 
\exp\lf(i\int d\si\hf\om_{ij}x^i\del_{\si}x^j
-\del_{\si}y_i(x^i+\th^{ij}\h{A}_j(x))\ri)\til{|y\ket} \cr
&=\int\D x\D y \exp\lf(i\int d\si\hf\om_{ij}(x^i-\th^{ik}y_k)\del_{\si}
(x^j-\th^{jl}y_l) \ri.\cr
&~~~~~~~~~~~~~~~~~~~~~~~~~~~~~~~~~~+\lf.\hf\th^{ij}y_i\del_{\si}y_j
-\del_{\si}y_i\th^{ij}\h{A}_j(x)\ri)\til{|y\ket} \cr
&=\int\D \xi\D y \exp\lf(i\int d\si\hf\om_{ij}\xi^i\del_{\si}\xi^j
-\hf\om_{ij}y^i\del_{\si}y^j
+\h{A}_i(\xi+y)\del_{\si}y^i\ri)\til{|y\ket}
}}
where $\xi^i=x^i-\th^{ij}y_j$ and $y^i=\th^{ij}y_j$.
For notational simplicity, we introduce the quantity
\eqn\defhW{
\h{W}(\h{A})=\lf\bra \exp\lf(i\int d\si \h{A}_i(y+\xi)\del_{\si}y^i\ri)
\ri\ket_{\xi},
}
where the expectation value is defined by
\eqn\defcorxi{
\bra \cdots\ket_{\xi}=\int \D\xi\,(\cdots)e^{i\int d\si\hf 
\om_{ij}\xi^i\del_{\si}\xi^j}.
}
Using $\h{W}(\h{A})$, eq.\phiinxi\ is written as
\eqn\phihW{
|\phi\ket_{{\rm NC}}
=\int\D y \;\h{W}(\h{A})\exp\lf(-i\int d\si\hf\om_{ij}y^i
\del_{\si}y^j\ri)\til{|y\ket}.
}

In the same way, $|\A\ket$ becomes
\eqn\Astatexi{
|\A\ket= \int\D y \;W(A)\exp\lf(-i\int d\si\hf\om_{ij}y^i
\del_{\si}y^j\ri)\til{|y\ket}
}
with
\eqn\defW{
W(A)=\lf\bra \exp\lf(i\int d\si A_i(y+\xi)\del_{\si}(y^i+\xi^i)\ri)
\ri\ket_{\xi}.
}
The equivalence of the two descriptions $|\phi\ket_{{\rm NC}}$ 
and $|\A\ket$ 
requires 
\eqn\map{
W(A)=\h{W}(\h{A}).
}
In the rest of this section, we will show that eq.\map\ 
gives the 
correct mapping between the ordinary gauge field $A$ and the
noncommutative gauge field $\h{A}$.

What we have to show are:
\item{(1)} $W(A)$ and $\h{W}(\h{A})$ have the same 
transformation property under the ordinary
and noncommutative gauge transformations,  respectively.
\item{(2)} Eq.\map\ has a nontrivial solution $\h{A}(A)$.

\noindent
If these two conditions are satisfied, the map obtained from \map\
agrees with the one defined in \SW.

First let us consider the condition $(1)$. 
The noncommutative gauge
transformation for $\h{A}$ is
\eqn\trfhA{
\cob \h{A}_i=\del_i\h{\la}+i\h{\la}\st\h{A}_i-i\h{A}_i\st\h{\la},
}
where the star product is defined by
\eqn\defstarp{
f\star g(x)=\sum_{n=0}^{\infty}{1\o n!}\lf({i\o2}\ri)^n\th^{k_1l_1}
\cdots\th^{k_nl_n}
\del_{k_1}\cdots\del_{k_n}f(x)\del_{l_1}\cdots\del_{l_n}g(x).
}
By expanding the exponential in $\h{W}(\h{A})$ and
performing the Wick contraction of $\xi$ using the two-point function
\eqn\corxi{
\big\bra \xi^i(\si)\xi^j(\si')\big\ket_{\xi}
=\big[(-i\om\del_{\si})^{-1}\big]^{ij} 
={i\o2}\th^{ij}\ep(\si-\si'),
}
$\h{W}(\h{A})$ can be  rewritten as
\eqn\exphW{\eqalign{
\h{W}(\h{A})&=1+\sum_{n=1}^{\infty}{i^n\o n!}\int d\si_1\cdots
d\si_n {\bf T}\big(\h{A}_{i_1}\st\cdots \st\h{A}_{i_n}\big)
\del_{\si_1}y^{i_1}\cdots \del_{\si_n}y^{i_n}\cr
&\equiv {\bf T}\exp_{\st}\Big[i\int 
d\si\h{A}_i(y)\del_{\si}y^i\Big]
}}
where ${\bf T}$ denotes  the time ordering. For example,
\eqn\exTord{
{\bf T}\Big\{f\big(y(\si_1)\big)\st g\big(y(\si_2)\big)\Big\}
=\th(\si_1-\si_2)f\st g+\th(\si_2-\si_1)g\st f.
}
Note that \exphW\ is the simplest example of the path integral
representation of the star product studied in \refs{\Kont,\CF}.
In \exphW, we generalized the notion of the star product to 
the product of functions at different points:
\eqn\genstar{
f(x)\st g(y)=\sum_{n=0}^{\infty}{1\o n!}\lf({i\o2}\ri)^n\th^{k_1l_1}
\cdots\th^{k_nl_n}
\del_{k_1}\cdots\del_{k_n}f(x)\del_{l_1}\cdots\del_{l_n}g(y).
}
From the expression \exphW, we can easily
see that
under the noncommutative gauge transformation \trfhA\
 $\h{W}(\h{A})$ transforms as
\eqn\hatWtrf{
\h{W}(\h{A}+\cob\h{A})=e^{i\h{\la}(y(\si_f))}\st\h{W}(\h{A})
\st e^{-i\h{\la}(y(\si_i))}
}
where we set the range of $\si$-integration to $[\si_i,\si_f]$.

Next we consider the transformation law of $W(A)$ under $\cob A_i=\del_i\la$. 
Naively, $W(A)$ transforms as 
\eqn\naivetrWA{\eqalign{
W(A+d\la)=&\lf\bra \exp\lf(i\int_{\si_i}^{\si_f} 
d\si A_i(y+\xi)\del_{\si}(y^i+\xi^i)\ri.\ri. \cr
&~~~~\qquad
+i\la\big(y(\si_f)+\xi(\si_f)\big)-i\la\big(y(\si_i)+\xi(\si_i)\big)
\Bigg)\Bigg\ket_{\xi} \cr
&=e^{i\h{\la}(\la,A)(\si_f)}\st W(A)\st e^{-i\h{\la}(\la,A)(\si_i)},
}}
where $\h{\la}(\la,A)$ is some function of $\la$ and $A$ whose explicit
form is not needed in the following discussion.
\naivetrWA\ shows that
 $W(A)$ has the same transformation property as that of $\h{W}(\h{A})$.
But we should take care of the divergence
coming from the contraction of $\xi^i$ and $\del_{\si}\xi^i$ with the
same argument, since $\bra\xi^i(\si_1)\del_{\si_2}\xi^j(\si_2)\ket$ 
is proportional to the delta function $\cob(\si_1-\si_2)$.
To make $W(A)$ finite, we regularize it by modifying the propagator of 
$\xi$:
\eqn\smearedW{
W(A)=\int \D\xi \,\exp\lf(i\int d\si \,\hf\om_{ij}\xi^iM(\del_{\si},\La)\xi^j
+A_i(y+\xi)\del_{\si}(y^i+\xi^i)\ri) 
}
with
\eqn\fexpansion{
M(\del_{\si},\La)=\del_{\si}-{\del_{\si}^3\o\La^2}.
} 
By the power counting, we can see that 
$\del_{\si}^3$ term in $M$ is sufficient to smear the
delta-function singularity in
$\bra\xi^i(\si_1)\del_{\si_2}\xi^j(\si_2)\ket$.
Explicitly, the propagator is given by
\eqn\expregprop{\eqalign{
\bra\xi^i(\si_1)\xi^j(\si_2)\ket&={i\o2}\th^{ij}\ep(\si_1-\si_2)
\Big(1-e^{-\La|\si_1-\si_2|}\Big) \cr
\bra\xi^i(\si_1)\del_{\si_2}\xi^j(\si_2)\ket&
=-i\th^{ij}{\La\o2}e^{-\La |\si_1-\si_2|}.
}}
This regularization does not break the gauge covariance of $W(A)$ \naivetrWA\ 
since we do not change the second term of the exponent in \smearedW. 

Now we consider the condition $(2)$.
For the equation $W(A)=\h{W}(\h{A})$ to have a solution, 
$W(A)$ has
to be the same form as $\h{W}(\h{A})$, namely 
the time ordered exponential. 
To explain that this is true, 
we expand $W(A)$ as
\eqn\wAandWn{
W(A)=\sum_{n=0}^{\infty}W_n(A)
}
where $W_0(A)=1$ and for $n>1$ $W_n(A)$ is defined by
\eqn\defWn{
W_n(A)=\lf\bra{1\o n!}\lf(i
\int d\si A_i(y+\xi)\del_{\si}(y^i+\xi^i)\ri)^n\ri\ket_{\xi}.
}
We write the first few terms of the
expansion of $W(A)$. $W_1(A)$ becomes
\eqn\Wone{
W_1(A)=\int d\si w_1(A)=\int d\si \Big[iA_i(y)\del_{\si}y^i
+w_1^{div}(A,\La)\Big]
}
with
\eqn\wonediv{
\lim_{\La\riya\infty}w_1^{div}(A,\La)=\hf\cob(0)\th^{kl}F_{kl}(y).
}
$W_2(A)$ becomes
\eqn\Wtwoform{
W_2(A)=\int d\si w_2(A)+{1\o2!}\int d \si_1d\si_2 {\bf T}\Big[
w_1\big(A(y(\si_1))\big)\st w_1\big(A(y(\si_2))\big)\Big]
}
where
\eqn\defwtwo{
w_2(A)=iA_i^{(2)}\del_{\si}y^i+w_2^{div}(A,\La)
}
and 
\foot{$A^{(2)}$ in this form was originally 
obtained by Garousi \refs{\Garo}.} 
\eqn\Atwowtwodiv{\eqalign{
A_i^{(2)}&=-\hf\th^{kl}(A_k,\del_lA_i+F_{li}), \cr
\lim_{\La\riya\infty}w_2^{div}(A,\La)&=-{1\o4}\cob(0)\th^{ij}\th^{kl}
F_{ik}F_{jl}.
}}
Here the bracket $(f,g)$ of two funcions $f$ and $g$
is defined by
\eqn\defsym{\eqalign{
(f,g)&\equiv\hf\int_{-1}^1dt \,f\exp\lf(t{i\o2}\overleftarrow{\del_i}\th^{ij}
\overrightarrow{\del_j}\ri)g \cr
&=\sum_{n=0}^{\infty}{1+(-1)^n\o2(n+1)!}
\lf({i\o2}\ri)^n\th^{i_1j_1}\cdots\th^{i_nj_n}\del_{i_1}\cdots\del_{i_n}f
\del_{j_1}\cdots\del_{j_n}g.
}}
In our formalism, this bracket appears as the
following correlation function:  
\eqn\braascor{
\Big(f(y(\si)),g(y(\si))\Big)
=\int d\si'\cob(\si'-\si)\Big\bra f(y(\si')+\xi(\si'))
g(y(\si)+\xi(\si))\Big\ket_{\xi}.
}
To obtain this relation, we used the formula
\eqn\cobepint{
\int d\si'\cob(\si'-\si)\ep(\si'-\si)^n={1+(-1)^n\o2(n+1)}.
} 

We can deduce the structure of $W_n(A)$
without the knowledge of the explicit form of it. 
Taking account of the combinatorial factors of the Wick contraction,
$W_n(A)$ becomes
\eqn\nthcont{
W_n(A)
=\sum_{\sum_ma_mk_m=n}\int {\bf T}\prod_m {1\o a_m!}
d\si_{i_1}\cdots d\si_{i_{a_m}}w_{k_m}\big(A(y(\si_{i_1}))\big)
\st\cdots\st w_{k_m}\big(A(y(\si_{i_{a_m}}))\big)
}
where the summation is taken over the partition of $n$, and $w_n(A)$
has the form
\eqn\wnform{
w_n(A)=iA_i^{(n)}(y)\del_{\si}y^i+w_n^{div}(A,\La).
}
$w_n^{div}(A,\La)$ is defined as the term in $w_n(A)$
which does not contain the factor 
$\del_{\si}y$. 
Notice that $A_i^{(n)}$ is finite
since the divergence in $w_n(A)$ comes from the contraction 
involving $n$ $\del_{\si}\xi$'s so the divergent term
does not have the factor $\del_{\si}y$. 

Eq.\nthcont\ means that $W(A)$ has the form of the time ordered
exponential, i.e.
\eqn\WAasTexp{
W(A)=\sum_{n=0}^{\infty}W_n(A)
={\bf T}\exp_{\st}\lf[\int d\si\sum_{n=1}^{\infty}
\Big(iA_i^{(n)}(y)\del_{\si}y^i+w_n^{div}(A,\La)\Big)\ri].
}
We can define the renormalized  Wilson loop $W^{fin}(A)$ by
\eqn\renormalizedWA{
W^{fin}(A)={\bf T}\exp_{\st}\lf[\int d\si
\sum_{n=1}^{\infty}iA_i^{(n)}(y)\del_{\si}y^i\ri].
}
This renormalization can be achieved by adding the counter term to
the path integral:
\eqn\defcount{\eqalign{
W^{fin}(A)=\int {\cal D}\xi \exp&\lf[\int d\si i\hf\om_{ij}\xi^iM(\del_{\si},\La)\xi^j
+iA_i(y+\xi)\del_{\si}(y^i+\xi^i)\ri. \cr
&~~~~-\lf.\sum_{n=1}^{\infty}
\Big(w_n^{div}(A(y+\xi),\La)+w_n'(A(y+\xi),\La)\Big)\ri].
}} 
$w_n'(A,\La)$ is the new counter term for the contraction
between $\del_{\si}\xi^i$ and $w_n^{div}(A,\La)$.

From \exphW\ and \renormalizedWA, 
the solution of $W^{fin}(A)=\h{W}(\h{A})$ is found to be
\eqn\hAinAn{
\h{A}=\sum_{n=1}^{\infty}A^{(n)}.
}
Here we list the first three terms of this expansion:
\eqn\Aonetothree{\eqalign{
A^{(1)}_i&=A_i~, \cr
A^{(2)}_i&=-\hf\th^{kl}(A_k,\del_lA_i+F_{li})~, \cr
A^{(3)}_i&=\hf\th^{kl}\th^{mn}A_k(\del_lA_m\del_nA_i
-\del_lF_{mi}A_n+F_{lm} F_{ni})+{\cal O}(\th^3)~.
}}
In principle, we can calculate
the right hand side of \hAinAn\ to any order in $\th$ by 
the simple rule of the Wick contraction.
It might be possible to write down the explicit
form of the map  to all order  in $\th$ in the same way as 
\refs{\Kont,\CF},
but we do not discuss it here.  

We comment on the relation between our result and the argument of
regularization in \SW. 
Since $\h{W}(\h{A})$ has the contraction between
$\xi$'s but does not contain the contraction of the form 
$\bra\xi(\si_1)\del_{\si}\xi(\si_2)\ket$, 
$\h{W}(\h{A})$ corresponds to the point-splitting regularization.
On the other hand, $W(A)$ is regularized so as to preserve the
covariance under the ordinary gauge symmetry. 
This is nothing but the mechanism of the appearance of
the two descriptions of gauge theory on $D$-branes due to the
possibility of two different regularizations. 
The only difference between our formalism and that in 
\SW\ is the dimension of the space on which the path integral is performed.
In \SW\ the path integral is defined on the whole 2-dimensional
worldsheet, while we consider that on the boundary of worldsheet.

\newsec{Discussions}
In this paper, we derived the map between $A$ and $\h{A}$ 
in a path integral form by comparing the boundary states in two different
gauges of the world-volume diffeomorphism.
To realize the noncommutative gauge symmetry, we chose the flat
measure $\D x$ in $|\phi\ket_{{\rm NC}}$. This measure does not respect 
the canonical transformation symmetry of $|\phi\ket$.
This difference between $|\phi\ket$ and $|\phi\ket_{{\rm NC}}$
may be related to the ``gauge equivalence of
the star product'' \Kont. 
This viewpoint deserves further study.

We comment on the generalization of our result to $U(N)$ gauge fields.
One natural way to generalize \main\ is to replace the exponential with
the trace of path ordered exponential. But 
the relation obtained by this
prescription is not enough to determine the map completely,  
since the gauge field has $N^2$ components.
To construct the complete map, we may have to use the additional
symmetry of the boundary state, such as the ``non-Abelian generalization
of diffeomorphism'' considered in \Ishin.

\vskip .2in

\centerline{{\bf Acknowledgments}}
I am grateful to N. Ishibashi and T. Kawano
for discussions and encouragement, and  K. Hayasaka,
A. Alekseev  and  A. Bytsko for pointing out a mistake in
the first version of this paper.  
I would also like to thank the organizers
of  the Summer Institute '99 in Fuji-Yoshida where this work started.
This work was supported in part by JSPS Research Fellowships for Young
Scientists.

\listrefs

\end